\documentclass{article} 
\usepackage{graphicx}
\begin{document}

\begin{center}
\Large\textbf{Interactions of MeV and GeV sterile neutrinos with matter. }
\end{center}

\begin{center}
\textbf{S. A. Alavi\footnote{E-mail: alavi@sttu.ac.ir, alialavi@fastmail.us}, T. Ebrahimi}\\

\textit{Department of Physics,  Sabzevar Tarbiat Moallem university
, P. O. Box 397, Sabzevar, Iran.}\\
\textit{}\\

 \end{center}

\emph{Both cosmology and astrophysics suggest the possible existence of sterile neutrino which is a dark matter candidate. The interaction of sterile neutrinos with matter in keV energy scale has been studied in the literature. In this paper we study the interaction of sterile neutrinos with atoms and  their role on ionization of atoms  in MeV and GeV energy scale. We also study the interaction of sterile neutrinos with nuclei in the MeV and GeV energy scale. We obtain the relevant cross sections for both these two interactions. Finally we compare our results with the results of keV energy range. }\\


\section{Introduction}
The discovery of neutrino masses suggests the likely existence of gauge singlet fermions (right-handed neutrinos) that participate in the neutrino mass generation [1]. A sterile neutrino  is a hypothetical neutrino that does  not interact via any of the fundamental  interactions of the Standard  Model except gravity. It is a right-handed neutrino  or a left-handed anti-neutrino. Such a particle belongs to a singlet representation with respect  to  the strong interaction and  the weak interaction and has zero weak hypercharge,  zero weak isospin and zero electric charge. Sterile neutrinos would still interact via gravity, so if they are heavy enough, they could explain cold dark matter or warm dark matter. The X-ray observations make use of the radiative decay of a sterile neutrino [2, 3],  can yield a non-negligible flux from concentrations of dark matter in astrophysical systems, such as, e.g., galaxies, clusters, and dwarf spheroidal galaxies [1, 4]. The photons emitted from decays sterile neutrinos can affect the formation of the first stars. Their production in a supernova can also explain the pulsar kicks and they have many other implications in astrophysics and cosmology. It is of interest, therefore to study  the interactions of sterile neutrinos in matter with the purpose of possibly using them to inform the direction of current and future experimental searches. \\

\section{Interaction with atoms and ionization of atom.}

In this section we study the scattering of a MeV sterile neutrino by an electron: $\nu_{s}e^{-}\rightarrow\nu_{e}e^{-}.$  
The relevant Feynman diagram depicted in Fig.1. The effective Hamiltonian for this scattering process is the same as ordinary neutrino-electron scattering and is given by [5]:
\begin{equation}
H_{eff}=\frac{G_{F}sin\theta}{\sqrt{2}}\bar{\nu_{e}}\gamma_{\mu}(1-\gamma_{5})\nu_{s}\bar{e}\gamma^{\mu}(c_{V}-c_{A}\gamma_{5})e
\end{equation}
Where: $c_{A}=\frac{-1}{2}$, $c_{V}=\frac{-1}{2}+2\sin^{2}\theta_{W}$, and $\sin^{2}\theta_{W}=0.23$ is the weak-mixing angle. To calculate the unpolarized cross section which means that no information about the electron spins is recorded in the experiment and to allow for scattering in all possible spin configurations, we make the following replacement:

\begin{equation}
\left|M\right|^{2}\rightarrow\overline{\left|M\right|^{2}}\equiv\frac{1}{(2s_{A}+1)(2s_{B}+1)}\sum_{all spins }\left|M\right|^{2}
\end{equation} 
\begin{figure}
\centering
\includegraphics[width=.5\textwidth]{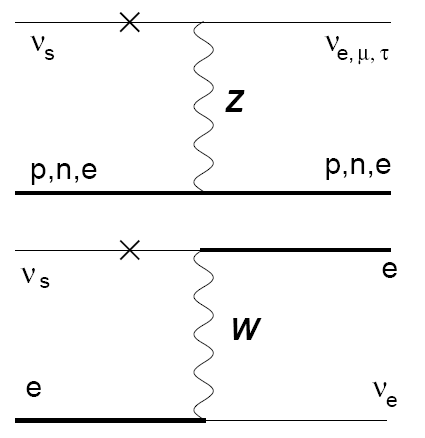}
\caption{ The Feynman diagrams describing interactions of sterile neutrinos in matter via charged and neutral currents.  }
\end{figure}
Where $s_{A}$, $s_{B}$ are the spins of the incoming particles. That is, we average over the spins of the incoming particles and we sum over the spins of the particles in the final state. Then we have:
\begin{eqnarray}
\frac{1}{2}\sum_{spins}\left|M\right|^{2}&=&4(G_{F}sin\theta)^{2}
((c_{V}+c_{A})^{2}(s-m_{e}^{2})
\times(s-m_{e}^{2}-m_{\nu_{s}}^{2})\nonumber
\\&+&(c_{V}-c_{A})^{2}(u-m_{e}^{2})\times(u-m_{e}^{2}-m_{\nu_{s}}^{2})\nonumber
\\&+&2(c_{V}^{2}-c_{A}^{2})m_{e}^{2}\times(t-m_{\nu_{s}}^{2}))
\end{eqnarray} 
\vspace{0.75 cm}

Where $s=(p_{\nu_{s}}+p_{e})^{2}$, $t=(p_{\nu_{e}}^{\prime}-p_{\nu_{s}})^{2}$, and $ u=(p_{e}-p_{\nu_{e}}^{\prime})^{2}$ are the Mandelstam variables. To calculate $\overline{\left|M\right|^{2}}$, we should first calculate the Mandelstam variables s, t, u. since $m_{\nu_{s}}\cong5k$eV [5], the sterile neutrinos  are relativistic which means that one can set $m_{\nu_{s}}\cong0$. Therefore the variables s, t, u take the following forms :
$s\approx m_{e}^{2}+2m_{e}E_{\nu_{s}}$, $t\approx-2E_{\nu_{e}}^{\prime}E_{\nu_{s}}(1-cos\theta_{\nu})$, $u\approx m_{e}^{2}-2m_{e}E_{\nu_{e}}^{\prime}$, 
Where $\theta _{\nu}$ is the scattering angle of the final neutrino $\nu_{e}$. we may write the differential cross section, in the symbolic form :

\begin{equation}
 d\sigma=\frac{\overline{\left|M\right|}^{2}}{F}dQ
\end{equation}
Where dQ is the invariant phase space factor and F is the initial flux.
By calculating $\overline{\left|M\right|}^{2}$, F and dQ, we obtain the following expression for the scattering cross section :

\begin{eqnarray}
 \sigma_{\nu_{s}e}&=&\int\frac{\overline{\left|M\right|}^{2}dQ}{4m_{e}E_{\nu_{s}}\upsilon}=
\frac{G_{F}^{2}sin^{2}\theta}{\pi\upsilon}
((c_{V}-c_{A})^{2}[2m_{e}(E_{e}^{\prime}-E_{\nu_{s}})+\frac{m_{e}^{4}}{E_{e}^{\prime}E_{\nu_{s}}}
\nonumber
\\&+&3m_{e}^{2}\frac{E_{e}^{\prime}}{E_{\nu_{s}}}-E_{e}^{'2}]
+c_{A}^{2}[2m_{e}(E_{e}^{\prime}-2E_{\nu_{s}})+3E_{\nu_{s}}^{2}]\nonumber
\\&+&(c_{V}E_{\nu_{s}})^{2}-2c_{V}c_{A}m_{e}E_{e}^{\prime})
\end{eqnarray}

 Where $\upsilon$ is the relative  velocity before scattering. \\
 Let us now study the interaction of a GeV sterile neutrino with an atom. Since the energy of the sterile neutrino is much more than the ionization energy and since the rest mass energy of the electron is 0.5 MeV,the final electron is relativistic, so we  can set $m_{e}\cong0$. The sterile neutrino is also relativistic in the GeV energy range.
The scattering cross section is as follows :
\begin{equation}
\sigma_{\nu_{s}e}=\frac{G_{F}^{2}sin^{2}\theta}{\pi\upsilon}\left[(c_{V}^{2}+3c_{A}^{2})E_{\nu_{s}}^{2}-(c_{V}-c_{A})^{2}E_{e}^{'2}\right].
\end{equation}
 
We now compare the result of ionization of atoms by keV and MeV-GeV sterile neutrinos. The momentum transfer to the electon for the keV sterile neutrino [5] is $\left|p'_{e}\right|=m_{\nu_{s}}$ and the electron kinetic energy in the final state is $T_{e}\cong\frac{m_{\nu_{s}}^{2}}{2m_{e}}=25$ eV, which is sufficient to ionize the atom.  In MeV-GeV case the momentum transfer to the electron is $\left|p'_{e}\right|\approx E_{\nu_{s}}$, so the electron kinetic energy in the final state for the MeV sterile neutrinos 
 is $T_{e}\approx \frac{E_{\nu_{s}}^{2}}{2 m_{e}}> 25$ eV and in the GeV case $T_{e}>> 25$ eV.
 In MeV case the energy is more than needed to ionize the atom and for the GeV sterile neutrinos it is much more than the ionization energy. Therefore if one can measure the electron spectrum resulting from these interactions it would peak at the energies more than or much more than 25 eV. As mentioned in [5] the well-defined prediction for the energy spectrum can be useful to reject potential backgrounds.\\
In MeV and GeV energy ranges the scattering of a sterile neutrino off the electrons in an atom 
 is not coherent. Because the momentum transfer is more than $m_{\nu_{s}}\approx 5$ keV, so it corresponds to the compton wavelengh smaller than the size of atom, $~ 10^{-8}$ cm, the sterile neutrino  does not scatter coherently of all the electrons in the atom, therefore we take $\sigma_{\nu_{s}A}\approx \sigma_{\nu_{s}e}$.\\

Let us compare the event rate of $\nu_{s}A$ scattering in keV and MeV-GeV energy ranges. It  is given by $R_{\nu_{s}A}=\sigma_{\nu_{s}A}\upsilon n_{\nu_{s}}N_{T}$, where $n_{\nu_{s}}$  is the number density of sterile neutrino dark matter, and $N_{T}$ is the number of target atom $A$ in the detector. As [5], we assume that the local mass density of dark matter is $0.4$ $GeV$ $cm^{-3}$, and it is only made of sterile neutrinos, their number density is $n_{\nu_{s}}= 8\times 10^{4} cm^{-3}(m_{\nu_{s}}/5 keV)^{-1}$. The number of target atoms is $N_{T}=(6\times 10^{29}/A)(M_{det}/ton)$, where $M_{det}$ is the mass of the detector. \\
It is shown that [1, 2] the life time of a sterile neutrino is proportional to the fifth power of its mass   $\Gamma_{\nu_{s}\rightarrow\nu_{a}\gamma} = \frac{1}{1.8 \times 10^{21}s}\sin^{2}\theta  m_{\nu_{s}}^{5}$. So the mixing angles for MeV-GeV sterile neutrinos are much smaller than the mixing angle of keV sterile neurinos. Taking $\left(\sin^{2}\theta\right)_{keV}=10^{-9}$ [5], we obtain the following values for MeV-GeV mixing angles :  $\left(\sin^{2}\theta\right)_{MeV}=10^{-24}$ and $\left(\sin^{2}\theta\right)_{GeV}=10^{-39}$.\\
To calculate the ratio of event rate of keV to MeV sterile neutrinos, we rewrite the MeV cross section (5) as follows :

\begin{eqnarray}
\sigma_{\nu_{s}e}&=&\frac{G_{F}^{2}sin^{2}\theta}{\pi\upsilon}E^{2}_{\nu_{s}}((c_{V}-c_{A})^{2}[2\frac{m_{e}}{E_{\nu_{s}}}(\frac{E_{e}^{\prime}}{E_{\nu_{s}}}-1)
\frac{m^{2}_{e}}{E^{2}_{\nu_{s}}}\frac{m_{e}^{2}}{E_{e}^{\prime}E_{\nu_{s}}} \nonumber
\\&+& 3\frac{m_{e}^{2}}{E_{\nu_{s}}^{2}}\frac{E_{e}^{\prime}}{E_{\nu_{s}}}
-\frac{E_{e}^{'2}}{E^{2}_{\nu_{s}}}]+c_{A}^{2}[2\frac{m_{e}}{E_{\nu_{s}}}(\frac{E_{e}^{\prime}}{E_{\nu_{s}}}-2)+3]+c_{V}^{2}   \nonumber
\\&-&2c_{V}c_{A}\frac{m_{e}}{E_{\nu_{s}}}\frac{E_{e}^{\prime}}{E_{\nu_{s}}})
\end{eqnarray}
Using  $c_{V}=-\frac{1}{2}+2\sin^{2}\theta_{W}=-0.04$, $c_{A}=-\frac{1}{2}=-0.5$ and taking  $E_{\nu_{s}}=100$ MeV, $\frac{E'_{e}}{E_{\nu_{s}}}=0.8$, we have :

\begin{equation}
\sigma^{MeV}_{\nu_{s}e}=\frac{G_{F}^{2}sin^{2}\theta}{\pi\upsilon}(0.62)E_{\nu_{s}}^{2}
\end{equation}

Therefore the desired ratio is given by :

\begin{equation}
R=\frac{R_{\nu_{s}A}^{keV}}{R_{\nu_{s}A}^{MeV}}=\frac{(0.75)Z^{2}\left(\sin^{2}\theta\right)_{keV}m^{2}_{\nu_{s}}}{(0.62)\left(\sin^{2}\theta\right)_{MeV}E^{2}_{\nu_{s}}}=1.9\times 10^{9}
\end{equation}
With the same procedure the ratio of event rate of keV to GeV sterile neutrinos is as follows :

\begin{equation}
R=\frac{R_{\nu_{s}A}^{keV}}{R_{\nu_{s}A}^{GeV}}=\frac{(0.75)Z^{2}\left(\sin^{2}\theta\right)_{keV}m^{2}_{\nu_{s}}}{(0.8)\left(\sin^{2}\theta\right)_{GeV} E^{2}_{\nu_{s}}}=1.9\times 10^{20}
\end{equation}

where we have chosen $E^{GeV}_{\nu_{s}}=10 GeV$ in Eqs.(10) and $Z=25$ in Eqs.(9) and (10).\\
 
\section{Spin flip of a nucleus.}

The sterile neutrinos are relativistic and the nucleus is nonrelativistic in the MeV energy range. If the spins of a nucleus are initially aligned in an external magnetic field, the flip of nuclear spins due to interaction (with sterile neutrinos) might be observed.
The target nuclei is nonrelativistic, so we take its four-component as $u(p_{N})=\sqrt{m_{N}}(\xi,\xi)^{T}$. The initial MeV sterile neutrino and the final neutrino are relativistic, so we take their spinors as: $u_{\nu_{e}}=\sqrt{2E'_{\nu_{e}}}(\xi_{\nu_{e}},0)^{T}$, $u_{\nu_{s}}=\sqrt{2E_{\nu_{s}}}(\xi_{\nu_{s}},0)^{T}$.
By choosing the z-axis in the direction of incident sterile neutrino, $\theta_{\nu}$ denotes the scattering angle of the final electron neutrino $\nu_{e}$. Therefore the two-components spinors of neutrinos are: $\xi_{\nu_{s}}=(0,1)^{T}$ and $\xi_{\nu_{e}}=(-sin(\frac{\theta_{\nu}}{2}),cos(\frac{\theta_{\nu}}{2}))^{T}$.
We assume the initial state nucleus has spin up along direction $(\theta_{s}, \phi_{s})$, where $\theta_{s}$ is the angle between z-axis and spin axis and $\phi_{s}$ is the azimuthal angle measured from the plane of scattering [6], so $\xi_{N}(\uparrow)=(cos(\frac{\theta_{s}}{2}), e^{i\phi_{s}}sin(\frac{\theta_{s}}{2}))^{T}$. The spin of the final-state nucleus will be down along the same direction, because the process is a spin flip process, so we take: $\xi_{N}(\downarrow)=(-e^{-i\phi_{s}}sin(\frac{\theta_{s}}{2}), cos(\frac{\theta_{s}}{2}))^{T}$.
Then the spin flip matrix element is as follows: 
\begin{eqnarray}
iM(\uparrow\rightarrow\downarrow)&=& -i\frac{G_{F}sin\theta}{\sqrt{2}}\bar{u}_{\nu_{e}}(p_{\nu_{e}}^{\prime})\gamma_{\mu}(1-\gamma_{5})u_{\nu_{s}}(p_{\nu_{s}})\nonumber
\\&\times& \bar{u}_{N}^{\downarrow}(p_{N}^{\prime})\gamma^{\mu}(c_{V}-c_{A}\gamma_{5})u_{N}^{\uparrow}(p_{N})= -i4\sqrt{2}G_{F}sin\theta c_{A}m_{N}\sqrt{E'_{\nu_{e}}E_{\nu_{s}}}\nonumber
\\&\times&(2sin\frac{\theta_{\nu}}{2}cos^{2}\frac{\theta_{s}}{2}-e^{i\phi_{s}}sin\theta_{s}cos\frac{\theta_{\nu}}{2}). 
\end{eqnarray}
 By averaging over the directions of the incident sterile neutrinos i.e. over the relevant angle between z-axis and spin $(\theta_{s},\phi_{s})$, the absolute value squared of $iM(\uparrow\rightarrow\downarrow)$, take the following form:
\begin{equation}
\overline{\left|M(\uparrow\rightarrow\downarrow)\right|^{2}}=32G_{F}^{2}sin^{2}\theta m_{N}^{2}c_{A}^{2}E'_{\nu_{e}}E_{\nu_{s}}(1-\frac{cos\theta_{\nu}}{3}).
 \end{equation}
We employ the same convention as the one in [5] and assume that, in an external magnetic field B, the spin up state is the ground state and the spin down state is the excited state. The energy difference between the upper and lower states is $2\mu_{N}B$ which is negligible in comparison with $m_{\nu_{s}}$ (which is in keV scale), in the expression $E_{\nu_{e}}\cong m_{\nu_{s}}-2\mu_{N}B$. Here $\mu_{N}$ is the magnetic moment of the nuclei, and we have assumed that the nuclei is infinitely heavy , because the energy of the sterile neutrino is in the MeV scale. The corresponding cross section is as follows : 
 \begin{eqnarray}
 \sigma(\uparrow\rightarrow\downarrow)&&= \frac{G_{F}^{2}sin^{2}\theta}{\pi\upsilon} c_{A}^{2}E_{\nu_{s}}E'_{\nu_{e}}(\int(1-\frac{cos\theta_{\nu}}{3})sin\theta_{\nu}d\theta_{\nu})\nonumber
\\&& =\frac{G_{F}^{2} sin^{2}\theta}{\pi\upsilon}2 c_{A}^{2} E_{\nu_{s}}^{2}. 
\end{eqnarray}

One can show that the cross section of opposite transition $\sigma(\downarrow\rightarrow\uparrow)$ is the same as $\sigma(\uparrow\rightarrow\downarrow)$. With the same calculations one can calculate the cross section of the transition $\sigma(\uparrow\rightarrow\uparrow)$ and $\sigma(\downarrow\rightarrow\downarrow)$ :
\begin{equation}
\sigma(\uparrow\rightarrow\uparrow)=\sigma(\downarrow\rightarrow\downarrow)=\frac{G^{2}_{F} sin^{2}\theta}{\pi\upsilon} \left(c^{2}_{A}+c^{2}_{V}\right) E^{2}_{\nu_{s}}.
\end{equation}

It is easy to show that the total cross section is given by :\\

$\sigma_{total}^{MeV}=\sigma(\uparrow\rightarrow\downarrow)+\sigma(\downarrow\rightarrow\uparrow)+\sigma(\uparrow\rightarrow\uparrow)+\sigma(\downarrow\rightarrow\downarrow)=$
\begin{equation}
2\frac{G^{2}_{F} sin^{2}\theta}{\pi\upsilon} \left(c^{2}_{V}+3c^{2}_{A}\right) E^{2}_{\nu_{s}}=2\frac{G_{F}^{2}sin^{2}\theta}{\pi\upsilon}(0.75)E_{\nu_{s}}^{2}.
\end{equation} 
Comparing with Eq.(8), we find that $\left(\sigma^{MeV}_{total}\right)_{spin flip}> \sigma^{MeV}_{\nu_{s}e}$.\\

It can be also shown that $\left(\sigma_{total}^{keV}\right)_{spin flip}>\left(\sigma_{total}^{MeV}\right)_{spin flip}$.\\
 
 Now, we study the scattering process of GeV sterile neutrinos  from a nucleus. In this case the nuclei get some kinetic energy after neutrino scattering, so its four-component spinor is of the form $u(p'_{N})=(\sqrt{p_{N}^{\prime}.\sigma}\xi(\downarrow), \sqrt{p_{N}^{\prime}.\overline{\sigma}}\xi(\downarrow))$ , where $p_{N}^{\prime}$ is the energy-momentum  four vector  of the nucleus after scattering and $\sigma$ is the pauli four matrices $\sigma=(1,\vec{\sigma})$ and $\overline{\sigma}=(1,-\vec{\sigma})$. The spin flip matrix element is :\\ 
\begin{eqnarray*}
iM(\uparrow\rightarrow\downarrow)&=&i4\sqrt{2}G_{F}sin\theta\sqrt{m_{N}E_{\nu_{s}}E'_{\nu_{e}}}sin\frac{\theta_{\nu}}{2}\nonumber
\\&\times&(c_{V}\left[cos^{2}\frac{\theta_{s}}{2}\left(\sqrt{E_{N}^{\prime}-p_{N}^{\prime}}-\sqrt{E_{N}^{\prime}+p_{N}^{\prime}}\right)\right]-c_{A}\nonumber
\\&\times&\left[\sqrt{E_{N}^{\prime}-p_{N}^{\prime}}e^{i\phi_{s}}sin\theta_{s}cot\frac{\theta_{\nu}}{2}
-\left(\sqrt{E'_{N}+p'_{N}}+\sqrt{E'_{N}-p'_{N}}\right)cos^{2}\frac{\theta_{s}}{2}\right])
\end{eqnarray*}

The  cross sections for different transitions are as follows :

$\sigma(\uparrow\rightarrow\downarrow)=\sigma(\downarrow\rightarrow\uparrow)=$
\begin{equation}
\frac{2}{3}\frac{G_{F}^{2}sin^{2}\theta}{\pi\upsilon}E_{\nu_{e}}^{'2}
\times\left[(c_{V}^{2}+2c_{A}^{2})+\frac{m_{N}}{E_{N}^{\prime}}(c_{A}^{2}-c_{V}^{2})-\frac{p_{N}^{\prime}}{E_{N}^{\prime}}(c_{A}^{2}-2c_{V}c_{A})\right]
\end{equation}
and\\
$\sigma(\uparrow\rightarrow\uparrow)=\sigma(\downarrow\rightarrow\downarrow)=$
\begin{equation}
\frac{2}{3}\frac{G_{F}^{2}sin^{2}\theta}{\pi\upsilon} E_{\nu_{e}}^{'2}
\left[(c_{A}^{2}+2c_{V}^{2})+\frac{m_{N}}{2E_{N}^{\prime}}(c_{A}^{2}-c_{V}^{2})-\frac{p_{N}^{\prime}}{E_{N}^{\prime}}(\frac{3}{2}c_{V}^{2}+\frac{1}{2}c_{A}^{2}-\frac{1}{2}c_{V} c_{A})\right]
\end{equation}
It is also straightforward to show that the total cross section in this case is as follows :

$\sigma_{total}^{GeV}=\sigma(\uparrow\rightarrow\downarrow)+\sigma(\downarrow\rightarrow\uparrow)+\sigma(\uparrow\rightarrow\uparrow)+\sigma(\downarrow\rightarrow\downarrow)=$
\begin{equation}
2\frac{G_{F}^{2}sin^{2}\theta}{\pi\upsilon}E_{\nu_{e}}^{'2}\left[2(c_{A}^{2}+c_{V}^{2})+\frac{m_{N}}{ E_{N}^{\prime}}(c_{A}^{2}-c_{V}^{2})-\frac{p_{N}^{\prime}}{E_{N}^{\prime}}(c_{V}^{2}+c_{A}^{2}-\frac{5}{3}c_{V} c_{A})\right]
\end{equation}
which in the limit $p_{N}^{\prime}\rightarrow 0$ gives Eq.(15) i.e., the total cross section for the Mev sterile neutrinos.\\ 
\section{conclusions}

Sterile neutrinos have important implications in cosmology and astrophysics which can be employed to inform the 
direction of current and future experimental searches. In this paper we have studied the interaction of sterile neutrinos 
with atoms and their role on ionization of atoms  in MeV and GeV energy scale. We have also studied the interaction of sterile neutrinos with nuclei in the MeV and GeV energy scale. We obtained the relevant cross sections for both these two interactions. we have compared our results with the results of keV energy range. Although it seems difficult to detect sterile neutrinos, but the experimental approach should not be hopeless in the long run.\\

\section{Acknowledgments}
We would like to thank A. Kusenko for his valuable comments.\\
                                     
\section{References.}
\noindent [1] A. Kusenko, hep-ph/0906.2968.\\
\noindent [2] P. B. Pal and L. Wolfenstein,  Phys. Rev. D25, 766 (1982).\\
\noindent [3]  V. D. Barger, R. J. N. Phillips, and S. Sarkar, Phys. Lett.
                B352, 365 (1995), hep-ph/9503295.\\
\noindent [4] K. Abazajian, G. M. Fuller, and M. Patel, Phys. Rev.
              D64, 023501 (2001),  astro-ph/0101524.\\
\noindent [5] S. Ando and A. Kusenko, Phys. Rev. D 81 (2010) 113006.\\
\noindent [6] M. E. Peskin and D. V. Schroeder, An Introduction to quantum field theory (Addison-Wesley, 1995).
\end{document}